\documentclass[showpacs,preprintnumbers,amsmath,pre]{revtex4}

\usepackage{graphicx}
\usepackage{dcolumn}
\usepackage{subfigure}
\usepackage{bm}

\begin{document}


\title{Confined granular packings: structure, stress, and forces}

\author{James W. Landry}
 \email{jwlandr@sandia.gov}
\author{Gary S. Grest}
 \affiliation{Sandia National Laboratories, Albuquerque, New Mexico 87185-1415}
\author{Leonardo E. Silbert}
 \affiliation{James Franck Institute, The University of Chicago, Chicago, IL 60637}
\author{Steven J. Plimpton}
 \affiliation{Sandia National Laboratories, Albuquerque, New Mexico 87185-0316}


\begin{abstract}
The structure and stresses of static granular packs in cylindrical
containers are studied using large-scale discrete element molecular
dynamics simulations in three dimensions.  We generate packings by both
pouring and sedimentation and examine how the final state depends on the
method of construction.  The vertical stress becomes depth-independent
for deep piles and we compare these stress depth-profiles to the
classical Janssen theory.  The majority of the tangential forces for
particle-wall contacts are found to be close to the Coulomb failure
criterion, in agreement with the theory of Janssen, while
particle-particle contacts in the bulk are far from the Coulomb
criterion.  In addition, we show that a linear hydrostatic-like region
at the top of the packings unexplained by the Janssen theory arises
because most of the particle-wall tangential forces in this region are
far from the Coulomb yield criterion.  The distributions of
particle-particle and particle-wall contact forces $P(f)$ exhibit
exponential-like decay at large forces in agreement with previous
studies.
\end{abstract}

\maketitle

\section{\label{sec:introduction}Introduction}

The formation and structure of granular packs has long been of interest
in both the engineering~\cite{Nedderman1992} and
physics~\cite{JaegerOct1996} communities.  One practical problem has
been how to characterize the behavior of granular materials in silos and
prevent silo failure.  A variety of simulation methods have been
developed to describe the stresses on the walls of a silo, though most
are confined to two dimensional ($2D$) systems.  Unfortunately, there is
wide disagreement as to the predictive power of these models and the
proper approach to take for accurate simulation~\cite{MassonApr2000,
HolstJan1999a, HolstJan1999b, SanadOct2001}.  Those simulations that are
carried out in three dimensions ($3D$) usually utilize finite-element
methods that provide little information on the internal structure or
forces in granular packs~\cite{ChenOct2001,GuinesOct2001}.  Most of the
recent $3D$ discrete-element simulations that have been performed employ
periodic boundary conditions in the two directions perpendicular to
gravity.  Though these studies provide useful information on the
internal structure of such packings~\cite{MakseMay2000,SilbertMar2002},
they give no information on vertical stresses or forces at the boundary.

The vertical stress in a silo has traditionally been described by the
pioneering 1895 theoretical work of Janssen~\cite{Janssen1895}.  This
analysis relies on treating a granular pack as a continuous medium where
a fraction $\kappa$ of vertical stress is converted to horizontal
stress.  The form of the vertical stress appears if one assumes that the
frictional forces between particles and walls are at the Coulomb failure
criterion: $F_t = \mu_w F_n$, where $F_t$ is the tangential friction
force, $F_n$ is the normal force at the wall, and $\mu_w$ is the
coefficient of friction for particle-wall contacts.  Numerous
improvements have been added over time, but in many cases their effect
on the theory is small~\cite{Nedderman1992}.  Recently, experiments have
been carried out on granular packs in silos to test the suitability of
Janssen's theory in ideal conditions.  These
studies~\cite{VanelMay1999,VanelFeb2000} measured the apparent mass at
the bottom of the silo as a function of the filling mass.  They found
the best agreement with a phenomenological theory containing elements of
Janssen's original model, which we describe in more detail in Sec IV.

We present here large-scale $3D$ discrete particle, molecular dynamics
simulations of granular packings in cylindrical containers (silos).  Our
aim is to understand the internal structure and vertical stress profiles
of these granular packings and reconcile our results with existing
theory.  A variety of methods simulating pouring and sedimentation are
used to generate the packings.  We show how the different methods of
filling the container affect the final bulk structure of the packings.
We evaluate the suitability of the Janssen theory to the observed
vertical stress profiles and test the validity of its assumptions.  We
show that the majority of particle-wall contact forces are close to the
Coulomb failure criteria, whereas particle-particle forces in the bulk
are far from yield.  Finally we show that the distribution of contact
forces in these packings show exponential-like tails, in the bulk, at
the side walls, and at the base~\cite{BlairApr2001,MuethMar1998}.

The simulation method is presented in Section I, where we also discuss
the various methods that were used to generate the packings.  In section
II, we show how the different methods affect the bulk structure of the
packings.  Section III presents the vertical stress profiles and
discusses their characteristics and we compare our results to the
classical theory of Janssen as well as two modified forms of the Janssen
analysis.  In Section IV we present our results on the distribution of
forces and test the Janssen prediction of Coulomb failure at the walls
of the cylinder.  We conclude and summarize the work in section V.

\section{\label{sec:simulation}Simulation Method}

We present molecular dynamics (MD) simulations in three dimensions on
model systems of $N$ mono-dispersed spheres of diameter $d$ and mass
$m$.  We vary $N$ from 20,000 to 200,000 particles.  The system is
constrained by a cylinder of radius $R$, centered on $x=y=0$, with its
axis along the vertical $z$ direction.  The cylinder is bounded below
with a flat base at $z=0$.  In some cases, a layer of randomly-arranged
immobilized particles approximately $2d$ high rests on top of the flat
base to provide a rough base.  The cylinders used vary in size from
$R=10d$ to $20d$.  This work builds on previous MD simulations of
packings with periodic boundary conditions in the $xy$
plane~\cite{SilbertMar2002}.

The spheres interact only on contact through a spring-dashpot
interaction in the normal and tangential directions to their lines of
centers.  Contacting spheres $i$ and $j$ positioned at $\mathbf{r}_i$
and $\mathbf{r}_j$ experience a relative normal compression $\delta =
|\mathbf{r}_{ij} - d|$, where $\mathbf{r}_{ij} = \mathbf{r}_i -
\mathbf{r}_j$, which results in a force
\begin{equation}
\mathbf{F}_{ij} = \mathbf{F}_n + \mathbf{F}_t.
\end{equation}
The normal and tangential contact forces are given by 
\begin{equation}
\mathbf{F}_{n} = f(\delta/d) (k_n \delta {\mathbf n}_{ij} - \frac{m}{2} \gamma_n \mathbf{v}_n)
\end{equation}
\begin{equation}
\mathbf{F}_{t} = f(\delta/d) (-k_t \mathbf{\Delta s}_t - \frac{m}{2} \gamma_t \mathbf{v}_t)
\end{equation}
where $\mathbf{n}_{ij} = \mathbf{r}_{ij}/r_{ij}$, with $r_{ij} =
|\mathbf{r}_{ij}|$. $\mathbf{v}_n$ and $\mathbf{v}_t$ are the normal and
tangential components of the relative surface velocity, and $k_{n,t}$
and $\gamma_{n,t}$ are elastic and viscoelastic constants,
respectively. $f(x) = 1$ for Hookean (linear) contacts while for
Hertzian contacts $f(x) = \sqrt{x}$.  $\mathbf{\Delta s}_t$ is the
elastic tangential displacement between spheres, obtained by integrating
tangential relative velocities during elastic deformation for the
lifetime of the contact.  The magnitude of $\mathbf{\Delta s}_t$ is
truncated as necessary to satisfy a local Coulomb yield criterion $F_t
\le \mu F_n$, where $F_t \equiv |\mathbf{F}_t|$ and $F_n \equiv
|\mathbf{F}_n|$ and $\mu$ is the particle-particle friction coefficient.
Frictionless spheres correspond to $\mu = 0$.  Particle-wall
interactions are treated identically, but the particle-wall friction
coefficient $\mu_w$ is set independently.  A more detailed description
of the model is available elsewhere~\cite{SilbertOct2001}.

Most of these simulations are run with a fixed set of parameters: $k_n =
2 \times 10^5 mg/d$, $k_t = \frac{2}{7} k_n$, and $\gamma_n =
50\sqrt{g/d}$.  For Hookean springs we set $\gamma_t = 0$.  For Hertzian
springs, $\gamma_t = \gamma_n$~\cite{PackingNote1}.  In these
simulations, it takes far longer to drain the energy out of granular
packs using the Hertzian force law, since the coefficient of restitution
$\epsilon$ is velocity-dependent~\cite{SchaeferJan1996} and goes to zero as the
velocity goes to zero.  We thus focused on Hookean contacts, which for
the above parameters give $\epsilon = 0.88$.  The convenient time
unit is $\tau = \sqrt{d/g}$, the time it takes a particle to fall its
radius from rest under gravity.  For this set of parameters, the
timestep $\delta t = 10^{-4} \tau$.  The particle-particle friction and
particle-wall friction are the same: $\mu = \mu_w = 0.5$, unless stated
otherwise.

All of our results will be given in dimensionless units based on $m$,
$d$, and $g$.  Physical experiments often use glass spheres of $d = 100
\mu m$ with $\rho = 2 \times 10^3 kg/m^3$.  In this case, the physical
elastic constant would be $k_{glass} \sim 10^{10} mg/d$.  A spring
constant this high would be prohibitively computationally expensive,
because the time step must have the form $\delta t \propto
k^{-\frac{1}{2}}$ for collisions to be modeled effectively.  We have
found that running simulations with larger $k$'s does not appreciatively
change the physical results~\cite{SilbertOct2001}.

We use a variety of techniques to generate our static packings.  In
method P1, we mimic the pouring of particles at a fixed height $Z$ into
the container. For computational efficiency a group of $M$ particles is
added to the simulation on a single timestep as if they had been added
one-by-one at random times.  This is done by inserting the $M$ particles
at non-overlapping positions within a thin cylindrical region of radius
$R-d$ that extends in $z$ from $Z$ to $Z-d$.  The $x$, $y$, and $z$
coordinates of the particles are chosen randomly within this insertion
region. The $x$, $y$, and $z$ coordinates of the particles are chosen
randomly within this insertion region.  The height of insertion $z$
determines the initial z-velocity $v_z$ of the particle --- $v_z$ is set
to the value it would have after falling from a height $Z$.  After a
time $\sqrt{2} \tau$, another group of M particles is inserted.  This
methodology generates a steady stream of particles, as if they were
poured continuously from a hopper (see Figure 1).  The rate of pouring
is controlled by setting $M$ to correspond to a desired volume fraction
of particles within the insertion region.  For example, for an initial
volume fraction of $\phi_i = 0.13$ and $R=10d$, the pouring rate is
$\approx 45$ particles/$\tau$.

Method P2 is similar, but the insertion region moves in $z$ with time,
so that the particles are inserted at roughly the same distance from the
top of the pile over the course of the simulation.  The insertion region
is the same as in method P1, with thickness $\delta z = 1$ and radius
$R-d$.  For the results presented here, the initial height is $10d$ and
the insertion point moves upward with velocity $v_{ins} = 0.15 d/\tau
\hat{z}$.  For 50,000 particles, the pouring region rises $150d$ over
the course of the simulation.  A 50,000 particle pack in a $R=10d$
cylinder is roughly $140d$ high, making this a reasonable rate for
pouring in particles at approximately the same height over a long run.
Different configurations were produced by using different random number
seeds to place the particles in the insertion region.  These two methods
are similar to the homogeneous ``raining'' methods used in
experiments~\cite{VanelNov1999}.

We also prepare packings that simulate particle sedimentation. In this
method non-overlapping particles with a packing fraction $\phi \approx
0.13$ are randomly placed in a cylindrical region of radius $R-d$
extending from $z = 10d$ to the top of the simulation box.  This tall,
dilute column of particles is then allowed to settle under the influence
of gravity in the presence of a viscous damping term -- each particle
$i$ feels an additional Stokes drag force $F^{damp}_i = -b v_i$, with
the damping coefficient $b = 0.20 m\sqrt{g/d}$.  The terminal velocity
$v_{term} = mg/b = 5\sqrt{dg}$ is the same velocity as that of a
free-falling particle that has fallen $25d/2$ from rest.  This method,
which we refer to as S2, closely approximates sedimentation in the
presence of a background fluid.  It also shares some similarities with
method P2, being very similar to pouring particles from a constant
height above the pile.  We also run the simulation with no viscous
damping, $b=0$, and refer to this as method S1. In both cases, we start
from the same initial configuration of particles but give the particles
different random initial velocities ranging from $-10 d/\tau$ to $10
d/\tau$ for the horizontal components and $-10 d/\tau$ to $0$ for the
vertical component to create different configurations.

In all cases, the simulations were run until the kinetic energy per
particle was less than $10^{-8} mgd$.  The resultant packing is
considered quiescent and used for further
analysis~\cite{SilbertMar2002}.  For method S1, the free-fall portion of
the simulation is a small fraction of the simulation time, with the
largest fraction of the simulation time devoted to dissipation of the
local vibrations of particles in contact.  For the other three methods,
the packs form as the pouring continues and lose their kinetic energy
very soon after the last particle settles on top of the pack.

These simulations were performed on a parallel cluster computer built
with DEC Alpha processors and Myrinet interconnects using a parallel
molecular dynamics code optimized for short-range interactions
\cite{SilbertOct2001,PlimptonMar1995}.  A typical simulation to create a
$50,000$ particle $R=10d$ packing through pouring takes $5 \times 10^6$
timesteps to complete and requires roughly $40$ CPU hours on $50$
processors.

\begin{figure*}
\includegraphics[width=1.0in]{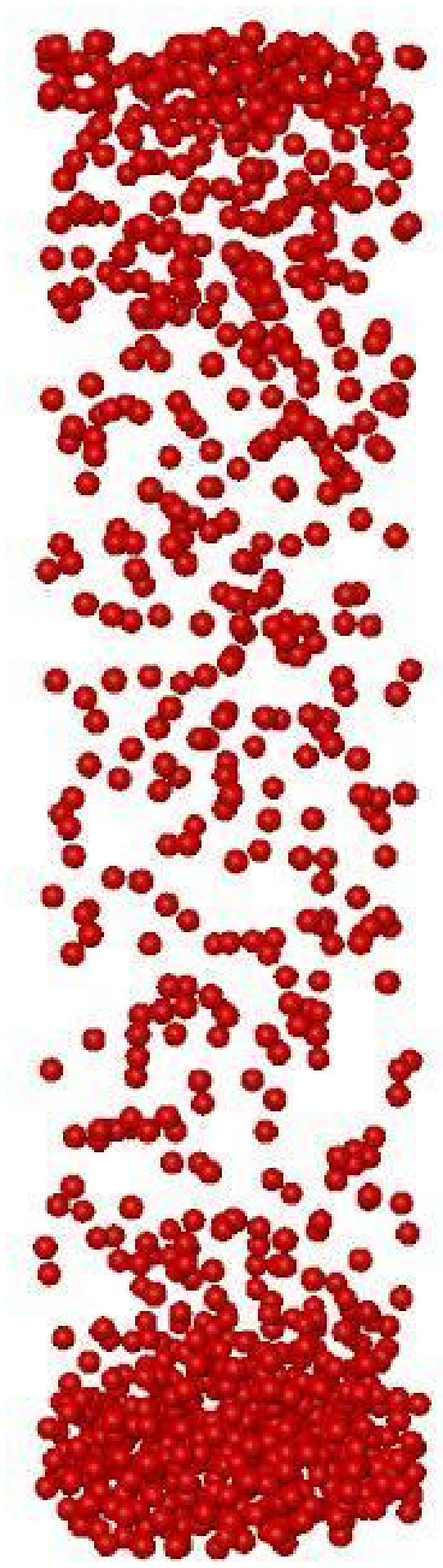}
\hspace{0.5in}
\includegraphics[width=1.0in]{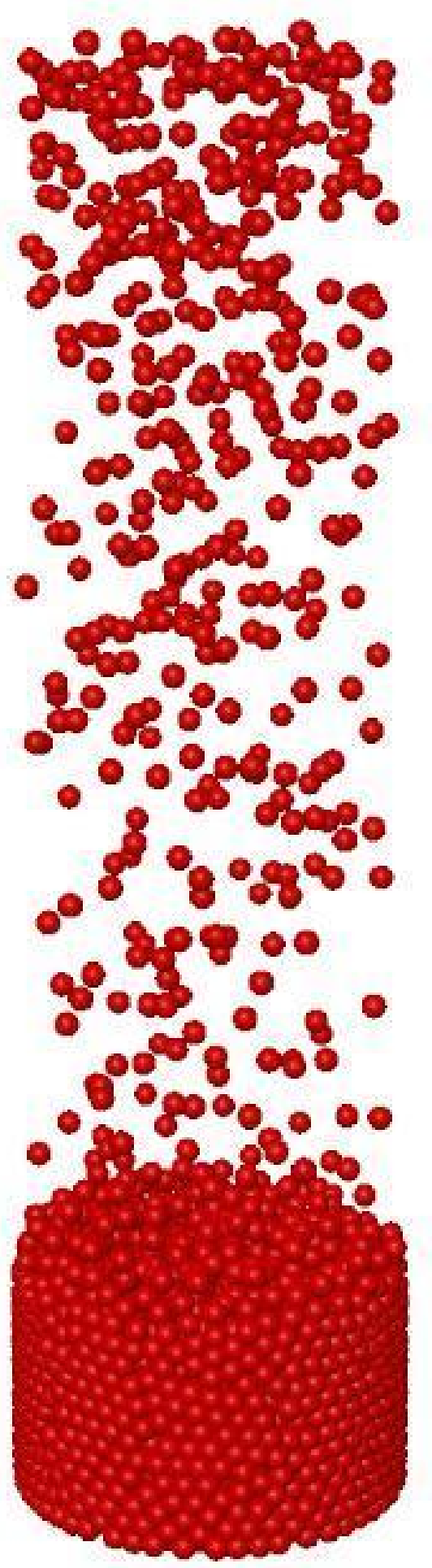}
\hspace{0.5in}
\includegraphics[width=1.0in]{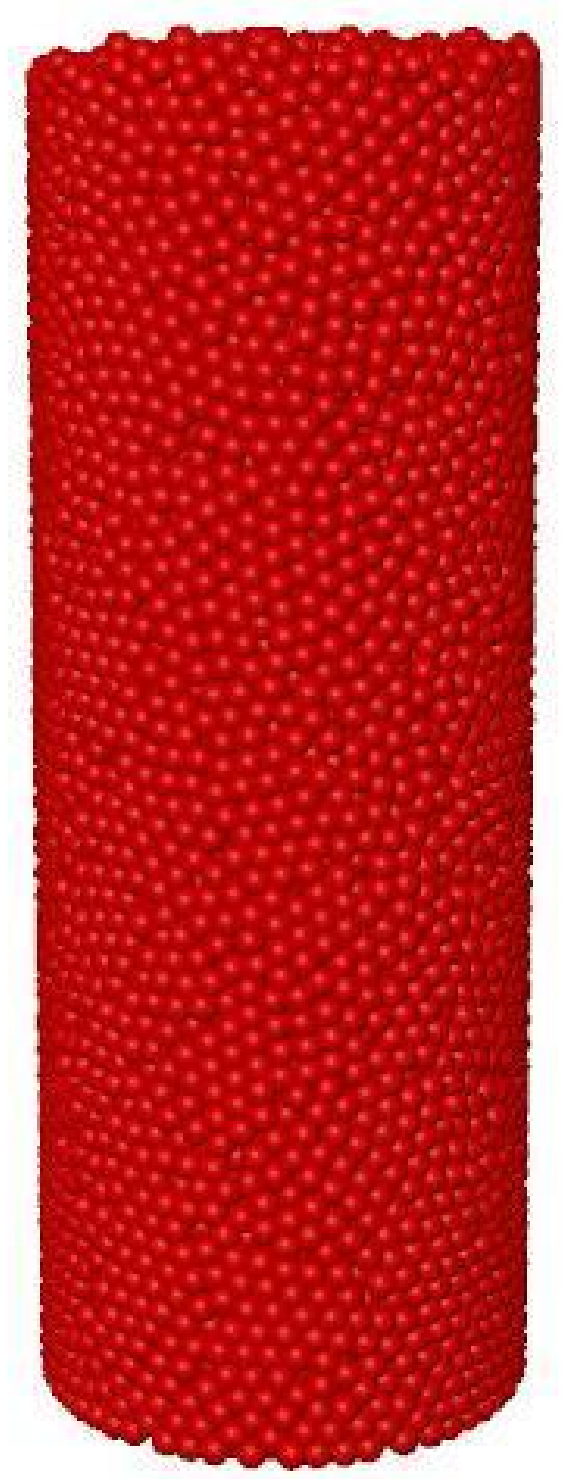}
\caption{\label{fig:pourmovie} Formation of a packing of $N = 20,000$
spheres in a cylindrical container of radius $10d$ onto a flat base.
The packing is constructed by pouring using method P1 from a height of
70d.  The configurations shown are for early, intermediate and late
times.  The final static pile has $\phi_f = 0.62$.}
\end{figure*}

\begin{figure*}
\includegraphics[width=1.0in]{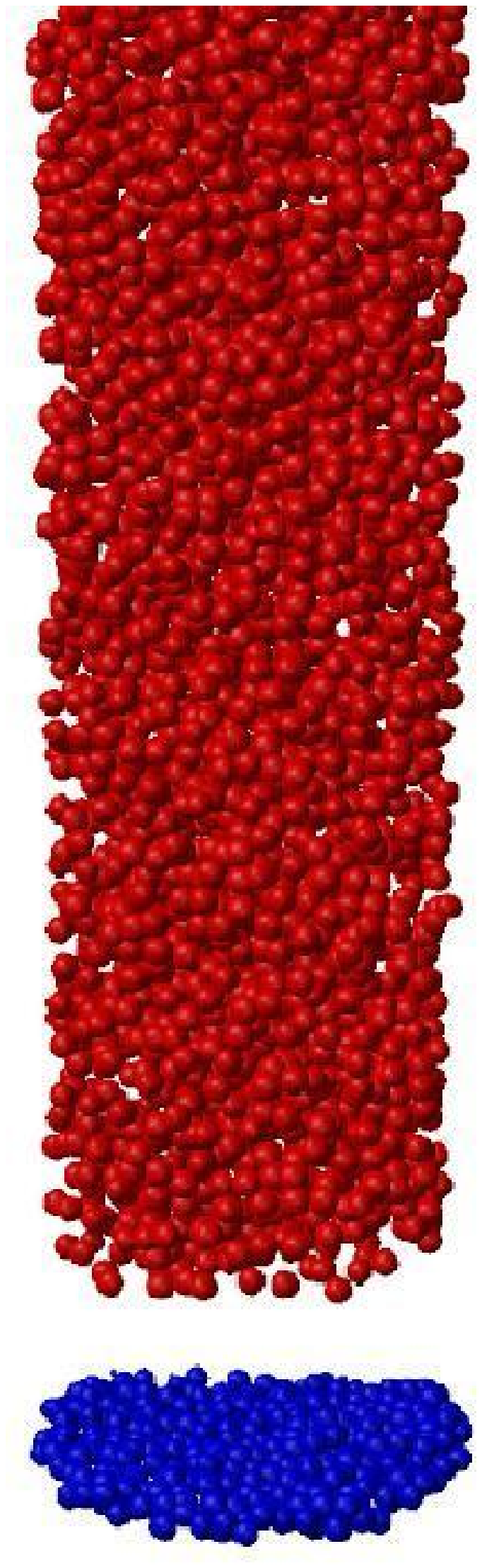}
\hspace{0.5in}
\includegraphics[width=1.0in]{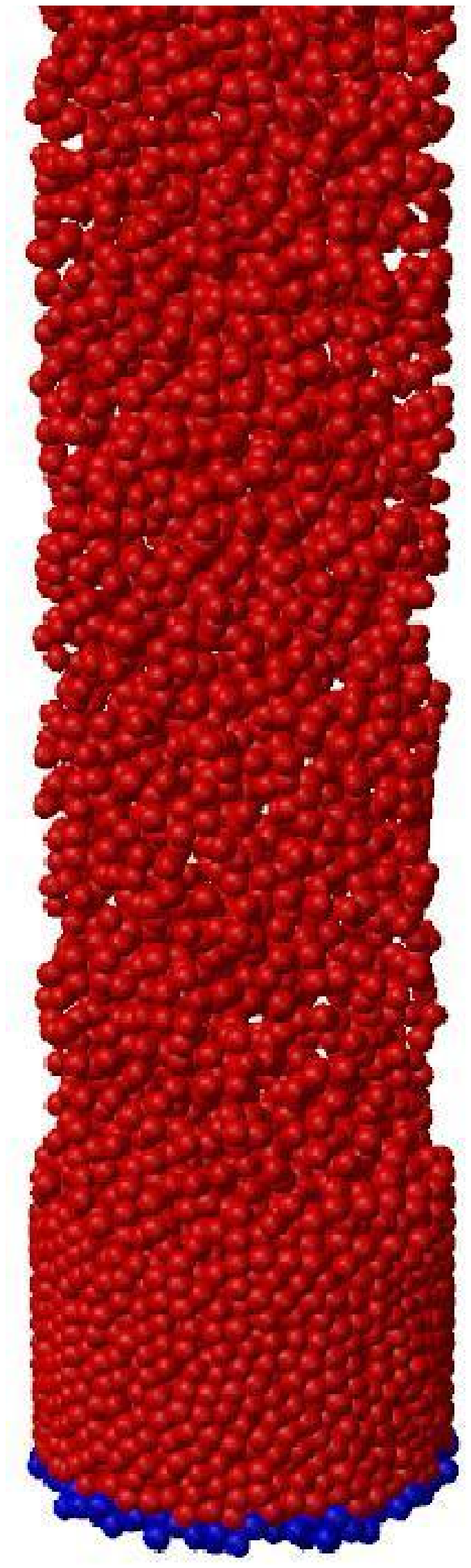}
\hspace{0.5in}
\includegraphics[width=1.0in]{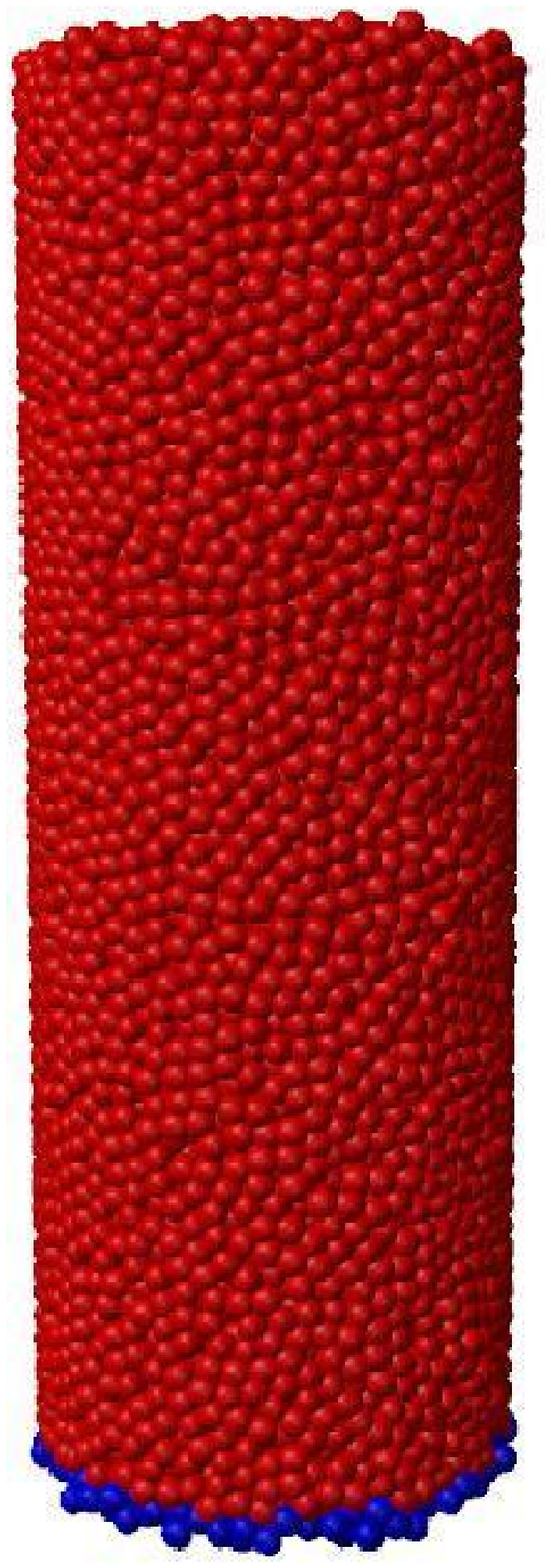}
\caption{\label{fig:sedmovie} Lower portion of the packing of $N = 20
000$ spheres in a cylindrical container of radius $R = 10d$.  The
packing is supported by a rough fixed bed (darker particles) and is
constructed by sedimentation using method S2.  The three configurations
shown are the initial configurations with volume fraction $\phi_i = 0.13$,
an intermediate one, and the final static pile with $\phi_f
\approx 0.60$.}
\end{figure*}

Figure~\ref{fig:pourmovie} shows a sample progression of our
simulations for method P1, while Figure~\ref{fig:sedmovie} shows
similar results for method S2, which are the two methods we focus on
in this paper. Both cases show a series of three snapshots over the
course of the formation of the pack~\cite{Raster,Merritt1997}.

\section{\label{sec:structure}Structure of the Packings}

The packings generated by these four methods had similar bulk
characteristics, though there were some differences in the final packing
fraction $\phi_f$ and coordination number $n_c$.  In all cases, the bulk
properties of the packings were the same for different random initial
conditions using the same method.  For a given set of initial conditions
such as pouring rate, pouring height or initial density, the height of
the resultant packing was the same to within $d/4$. The resulting
packing fraction $\phi$ and coordination number $n_c$ within the pack
were reproducible for a given set of initial conditions.  Because of
this, we frequently averaged over multiple runs with different random
initial conditions to improve statistics in the presentation that
follows.

\begin{figure}
\includegraphics[width=2.25in,angle=270,clip]{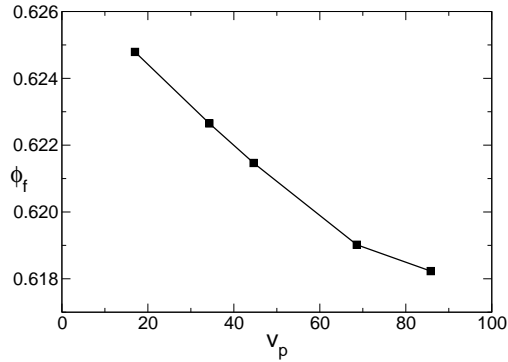}
\caption{\label{fig:pour_rate}Final average packing fraction $\phi_f$ as
a function of pouring rate $v_p$ (in units of $1/\tau$).  Results are
for packings of $50,000$ particles with $R=10d$ poured from a height of
$180d$ with method P1.  The line is  a guide to the eye.  Slower
pour rates create denser packings.}
\end{figure}

Small differences in the physical structure of the packs were observed
that depend slightly on the generation method.  In general, packings
created by pouring were denser than those created by sedimentation.  For
otherwise identical $50,000$ particle packings in a cylinder of radius
$R=10d$ with default parameters, packings created using methods P1 had
an average volume fraction $\phi_f \approx 0.621$ and for P2 had an
average volume fraction of $\phi_f \approx 0.614$ using a pouring rate
of 45 particles$/\tau$. Those created using methods S1 had an average
volume fraction of $\phi_f \approx 0.597$ and those using method S2 had
an average volume fraction of $\phi_f \approx 0.594$.  These differences
were reproducible over different initial conditions.  The difference
between pouring and sedimentation seems to arise from the much longer
times involved in pouring, because the energies involved in both methods
are not dissimilar.  The longer time scales required to form packs
through pouring seem to allow particles more time to settle and
rearrange, thus creating denser packs.  Sedimentation occurs over much
faster time scales and seems to lock the particles into metastable
configurations that are less dense.  For method P1, increasing the
height from which the particles were poured also increased the density
of the final pack, though the effect was slight.  This effect probably
arises from the greater kinetic energy of the particles when they hit
the pack, which allows them to explore more phase space, resulting in
denser packs.  The pouring rate also affects the final density $\phi_f$,
with faster pouring rates producing looser packings as shown in
Figure~\ref{fig:pour_rate}.  This is the same effect as above, with
faster pouring rates forcing particles into looser meta-stable
configurations.  The final packing fraction $\phi_f$'s for Method P2 are
consistently lower than those for method P1.  This is due to the change
in kinetic energy, because the kinetic energy of pouring particles in
method P2 is much smaller than in P1. As was reported earlier for
periodic systems~\cite{SilbertMar2002}, more dilute initial packing
fractions $\phi_i$ result in larger final packing fractions $\phi_f$,
and we see this behavior also for our simulations using method S1.  This
is the same effect as increasing the pouring height, because more dilute
columns with smaller $\phi_i$ are also taller and thus have greater
potential energy.  In model S2 the final velocity of the falling
particles is limited by the drag to a small terminal velocity.  This
removes any excess kinetic energy and the final packing fractions of
these packings are independent of the initial state.  Finally, the force
law chosen also has a very slight effect on the final structure of the
pack.  Replacing the Hookean force law with Hertzian results in a
slightly denser pack.  We thus affirm the history-dependence of granular
packings: the structure of the resultant packing is dependent on the
particular method used to generate it~\cite{VanelNov1999}.

\begin{figure}
\includegraphics[width=2.25in,angle=270,clip]{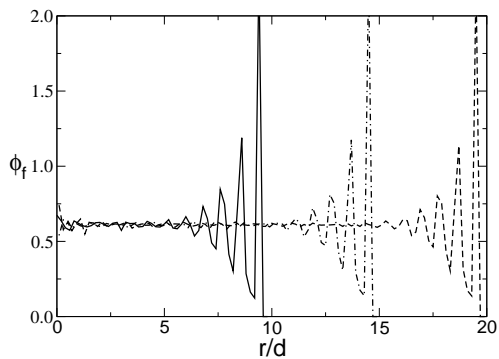}
\caption{\label{fig:rad-den}Final volume fraction $\phi_f$ of packings
as a function of radius for packings of $N = 50,000$ and $R=10$,
$N=82,000$ and $R=15$, and $N=144,000$ and $R=20$ using method S2.  The
effects of the wall penetrate about $4d$ in each case.}
\end{figure}

We find that significant particle ordering is seen at the cylinder
walls, but this boundary effect penetrates only a few diameters into
the bulk for cylinders of various radii. Figure~\ref{fig:rad-den}
shows the final packing fraction as a function of radius for a set of
packings created using the same parameters in cylinders of different
radii using method S2. In all these cases $\phi_f$ quickly approaches
the bulk value irrespective of the size of the container.  In
addition, the decay length $\nu$ is independent of size and extends over
$\nu \approx 4d$ for all $R$.

Previous studies of granular packings have been concerned with the
stability of
packings~\cite{MasonSep1997,MakseMay2000,OhernFeb2002,SilbertMay2002}.
The stability of a packing is based on the average number of contacts
per particle -- the coordination number $n_c$.  The theoretical
limit for stability for particles with friction is $n_c =
4$~\cite{AlexanderMar1998}.  Packings with $n_c = 4$ are said to be
isostatic, while those with $n_c > 4$ are hyperstatic - they have more
contacts than are needed for mechanical stability.  A previous
study~\cite{SilbertMay2002} of packings with horizontal periodic
boundary conditions using the same model concluded that frictional
packings are always hyperstatic.  Using methods S1 and S2, we see
identical results for $\phi_f$ and $n_c$ to those previous measurements
in the inner core of our packings for particles more than $5d$ from the
outer wall, which should remove any ordering effects originating from
the wall.  Packings generated by methods P1 and P2 are also hyperstatic.
This suggests that the previous conclusions of hyperstaticity also apply
in the bulk of silos and that the walls have only a small effect on the
physical structure of packings.  The method used to create the packings
seems to have a much larger effect.

\section{\label{sec:stresses}Distribution of Stresses}

Of particular interest in the construction of silos is the distribution
of stresses in a cylindrical packing~\cite{Nedderman1992}.  In a liquid,
hydrostatic pressure increases with depth.  Granular materials support
shear, so the side walls of a container can support some of this
pressure.  The problem of the resultant vertical stress in a silo after
filling has a long history, beginning with Janssen in 1895.  Janssen's
analysis\cite{Janssen1895,Rayleigh1906} of the stress in a silo rested
on three assumptions: the granular particles are treated as a continuous
medium, a vertical stress $\sigma_{zz}$ applied to the material
automatically generates a horizontal stress $\sigma_h = \kappa
\sigma_{zz}$, and the frictional forces between particles and the wall
are at the point of Coulomb failure ($F_t = \mu_w F_n$), where the
frictional force can no longer resist tangential motion of the particle
and have a specific direction.  In our case, this direction is upward as
the particles settle.  Using our simulations we can test some of these
assumptions.

For a cylindrical container of radius $R$ with static wall friction
$\mu_w$ and granular pack of total height $z_0$, the Janssen analysis
predicts the vertical stress $\sigma_{zz}(z)$ at a height $z$ is
\begin{equation}
\sigma_{zz}(z) = \rho g \bar{l}\left[1 -
\exp\left(-\frac{z_0 - z}{l}\right)\right]
\label{eq:janssen}
\end{equation}
where the decay length is $l = \frac{R}{2\kappa\mu_w}$. $\kappa$
represents the fraction of the weight carried by the side walls, $\rho$
is the volumetric density, and $z_0$ is the top of the packing.  In our
case, $\rho = \phi_f \rho_p$, where $\rho_p = 6m/\pi d^3$ is the density
of a single particle.  Standard Janssen analysis mandates that $l=\bar{l}$, so
that $l$ is the only free parameter.  As seen below in
Figure~\ref{fig:pz-janssen}, this single parameter formula does not
provide a good qualitative fit to our data.  We have generalized the
formula to include a two parameter fit with $l \ne \bar{l}$.  This
separates the asymptote from the decay length.  This generalization is
similar to the one proposed by Walker to address the experimental fact
that stresses are not uniform across horizontal slices, as was assumed
in the original Janssen analysis~\cite{Walker1966,Walters1973}.

Another two-parameter fit was proposed by Vanel and
Cl\'{e}ment~\cite{VanelMay1999} to reconcile their experimental findings
with Janssen theory.  The fit assumes a region of perfect
hydrostaticity, followed by a region that conforms to the Janssen
theory.  
\begin{eqnarray}
z_0 - z < a & : & \sigma_{zz}(z) = \rho g (z_0 - z) \nonumber\\ 
z_0 - z > a & : & \sigma_{zz}(z) = \rho g \left(a + l\left[1 - \exp\left(-\frac{z_0 - z - a}{l}\right)\right]\right)
\label{eq:twoparm}
\end{eqnarray}
This hydrostatic region is also predicted by a model of Evesque
and de Gennes~\cite{EvesqueNov1998}.

\begin{figure}
\includegraphics[width=2.25in,angle=270,clip]{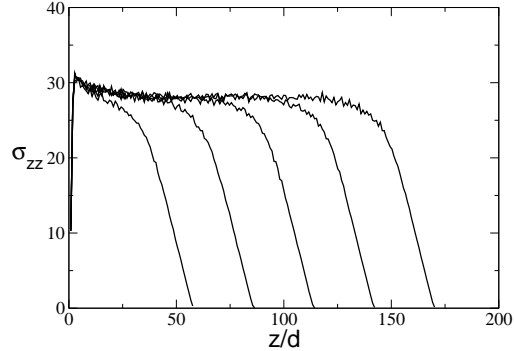}
\caption{\label{fig:pz-diffheight}Vertical stress $\sigma_{zz}$ in units
of $mg/d^2$ for $N = 20000$ to $60000$ packings with a rough base, using
$\mu = \mu_w = 0.5$ and $R=10d$ for method S2.  Data for each value of
$N$ is averaged over 6 runs.}
\end{figure}

Vertical stress profiles of packings for different numbers of particles
using method S2 are shown in Figure~\ref{fig:pz-diffheight}.  As the
height of the packing increases, the region of height-independent stress
also increases.  We estimate that a ratio of height to radius of $h/R
\approx 6$ is required to see this behavior, though this may be somewhat
dependent on our cylindrical geometry and also the dimensionality of the
system, since this ratio is smaller than that observed in
2D~\cite{MassonApr2000,LandryNov2002}.  There is a slight increase in
the vertical stress at the base of all of these packings.  This is a
generic feature of our packings, visible in packings with rough and flat
bases, and is a boundary effect at the base.  We ignore this small
region in our subsequent analyses.

We show a fit of the $N=50,000$ stress profile to the Janssen formula
Eq.~\ref{eq:janssen} in Figure~\ref{fig:pz-janssen}a.  We obtain the fit
by setting the asymptote $\rho g \bar{l}$ equal to the value of the
stress in the height-independent region.  This section is independent of
depth and thus is the controlling factor for the Janssen fit.  We used
the standard $\chi^2$ measure of goodness of fit to evaluate the fit,
where $\chi^2 = \sum_{i=1}^N \frac{(y_i - x_i)^2}{N-1}$, $N$ is the
number of data points, $x_i$ is the simulation data, and $y_i$ are the
points from the fit.  In this and subsequent fits, we do not use the
bottom $25d$ of the cylinder, as the uptick of the stress there is a
boundary effect.  All fit parameters are summarized in
Table~\ref{tab:stress-fits}.  The Janssen fit is relatively poor
($\chi^2= 10.5$), and it substantially under-predicts the stress in the
turnover region.  As in the experimental data by Vanel and
Cl\'{e}ment~\cite{VanelMay1999}, the hydrostatic region is larger than
predicted by the standard Janssen analysis.  We also fit our stress
profile to the modified Janssen form $(l \ne \bar{l})$, taking $\bar{l}$
from the asymptote as before and fitting $l$ as a free parameter.  This
fit is better ($\chi^2 = 1.03$).  However, this form also under-predicts
the size of the linear region and overshoots the data for large $z$, as
shown in Figure~\ref{fig:pz-janssen}b.  As the stress increases linearly
with $z$ near the top of the packing, it is not surprising that the best
fit was obtained with the two-parameter Vanel-Cl\'{e}ment form,
Eq.~\ref{eq:twoparm}, with a $\chi^2 = 0.092$.  These results are
qualitatively in agreement with the results obtained by Vanel and
Cl\'{e}ment: we obtain $\kappa$'s greater than $1$ for the two-parameter
fit and $\kappa$'s smaller than one for the standard Janssen fit.  It is
difficult to provide a direct prediction for the value of $\kappa$ we
expect~\cite{Nedderman1992}.  The latter two fits (modified Janssen and
the two-region fit) do not have a theoretical basis, but clearly
represent the data much better.  There is a substantial region of linear
hydrostatic pressure at the top of the packing that both the classical
and modified Janssen theory do not account for.

\begin{figure}
\includegraphics[width=2.25in,angle=270,clip]{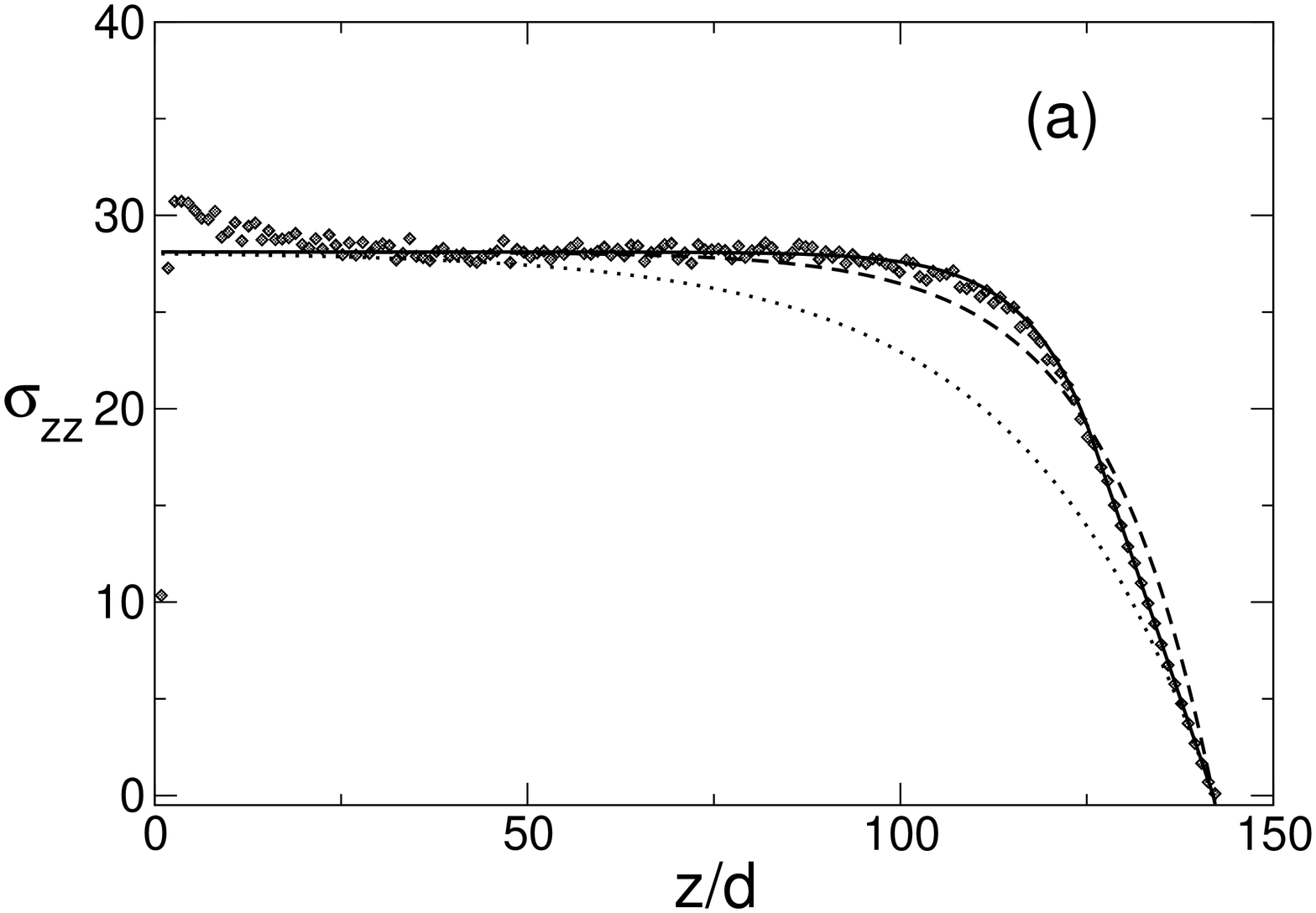}
\includegraphics[width=2.25in,angle=270,clip]{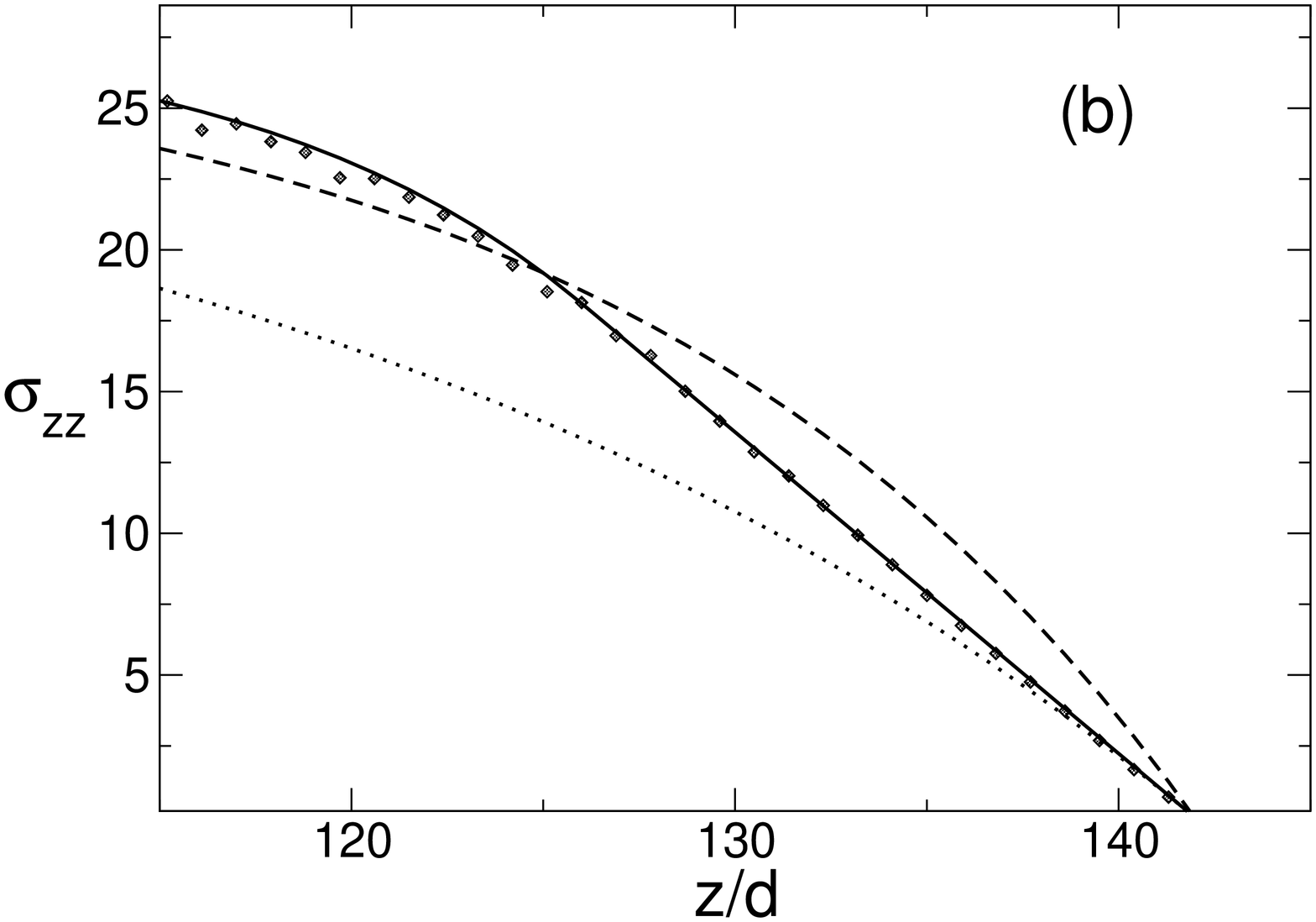}
\caption{\label{fig:pz-janssen}Vertical stress $\sigma_{zz}$ in units of
$mg/d^2$ for $N = 50000$ using method S2.  The data is represented by
the diamonds.  The dotted line is a fit to the Janssen expression with
$l=\bar{l}$, Eq.~\ref{eq:janssen}.  The dashed line is a fit to the
modified Janssen expression with $l \ne \bar{l}$.  The solid line is a
fit to the two parameter theory, Eq.~\ref{eq:twoparm}. (b) is a blowup
of the turnover region on the right side of (a).}
\end{figure}

\begin{table}
\caption{\label{tab:stress-fits} Results of the fits for vertical stress in
  packings using method S2 and the corresponding physical parameters.}
\begin{ruledtabular}
\begin{tabular}{cccc}
Packing Friction & Janssen & modified Janssen & Vanel-Cl\'{e}ment \\ \hline
$\mu = 0.5$ & $\chi^2 = 10.5$ & $\chi^2 = 1.03$ & $\chi^2 = 0.092$ \\
$\mu_w = 0.5$ & $\kappa = 0.404$ &  $\kappa = 0.677$ & $\kappa = 1.14$ \\
 & $l/d = 24.8$ & $l/d = 14.8$ & $l/d = 8.76$ \\
 &  & $\bar{l}/d = 24.8$ & $a/d = 16.0$ \\
$\mu = 0.5$ & & $\kappa = 0.168$ & $\kappa = 0.218$ \\
$\mu_w = 2.0$ & & $l/d = 14.9$ & $l/d = 11.5$ \\
 & & $\bar{l}/d = 22.7$ & $a/d = 11.5$ \\
\end{tabular}
\end{ruledtabular}
\end{table}

We find similar results for all other methods except S1.  Method S1 is
somewhat unphysical, since the particles hit the packing with increasing
kinetic energy as the simulation progresses.  The vertical stress we
observe in this case is substantially larger than that observed for
other methods and is noticeably peaked near the top of the sample.  This
arises because the large velocities of accelerating particles
excessively compact the pack at impact.  The pack then attempts to
relax, but the side walls exert their own pressure on the pack, keeping
it in its ``stressed'' position, yielding a total pressure greater than
hydrostatic and freezing in this kinetic stress.  Although there is a
large difference in the stress profiles between packings generated by
method S1 and S2, $\phi_f$ of the former is only slightly larger.

\begin{figure}
\includegraphics[width=2.25in,angle=270,clip]{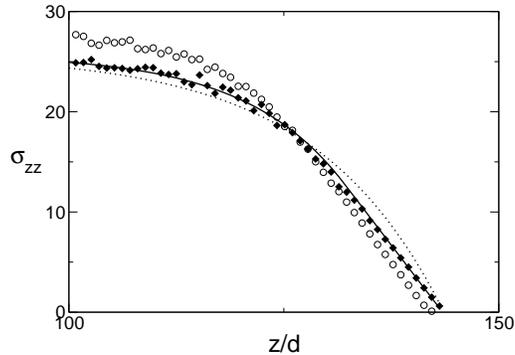}
\caption{\label{fig:pz-highfric}Vertical stress $\sigma_{zz}$ in the top
part of $N = 50000$ packings, with $\mu_w = 2$ (diamonds) and $\mu_w =
0.5$ (open circles), both using method S2.  The dotted line is a fit to
the $\mu_w = 2$ data with the modified Janssen formula with $l \ne
\bar{l}$ and the straight line is a fit to the same with the
two-parameter Vanel-Cl\'{e}ment formula.}
\end{figure}

To test the underlying assumptions of the Janssen analysis, we varied
the particle-wall friction $\mu_w$. First we set $\mu_w = 0$, which
removed any particle-wall friction.  This prevents the side walls from
supporting any weight and is similar to unconfined packings.  The result
is a vertical stress that increases linearly with height, exactly as in
the hydrostatic case and as expected from the Janssen analysis.  Another
test was to increase the particle-wall friction, setting $\mu_w = 2.0$.
This ensures a very high limit for the Coulomb failure criterion.  We
compare the stress profile of the $\mu_w = 2.0$ case to our standard
$\mu_w = 0.5$ case in Figure~\ref{fig:pz-highfric}, both with $\mu =
0.5$.  The higher wall-friction case has a lower height-independent
stress, because the larger the $\mu_w$, the more the walls can support.
However, this difference is not large, and using $\mu_w = 2.0$ to obtain
$\kappa$ values results in unreasonably low values, as seen in
Table~\ref{tab:stress-fits}.  The modified Janssen from gives $\kappa =
0.168$, and the two-parameter fit gives $\kappa = 0.218$.  $\kappa$
should be a feature of the material used and not vary greatly when the
wall friction is changed~\cite{Nedderman1992}.  All of these fits use
part of the Janssen theory, and the discrepancy in $\kappa$ arises
because the third assumption of the Janssen analysis is not satisfied:
the tangential forces at the wall for the $\mu_w = 2.0$ case are
considerably less than $\mu_w F_n$ and thus far from the Coulomb failure
criterion, as seen in Sec. V.

\begin{figure}
\includegraphics[width=2.25in,angle=270,clip]{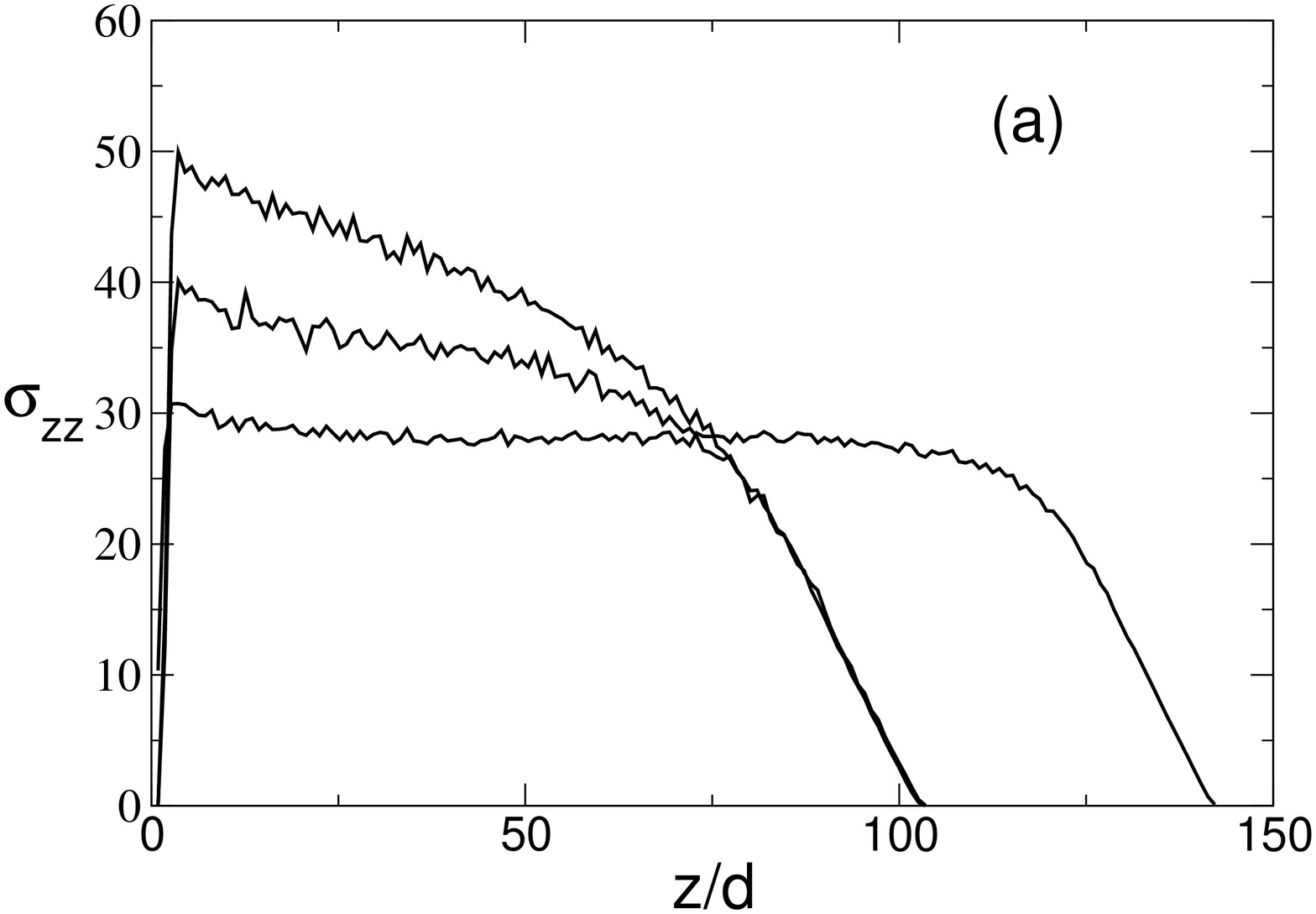}
\includegraphics[width=2.25in,angle=270,clip]{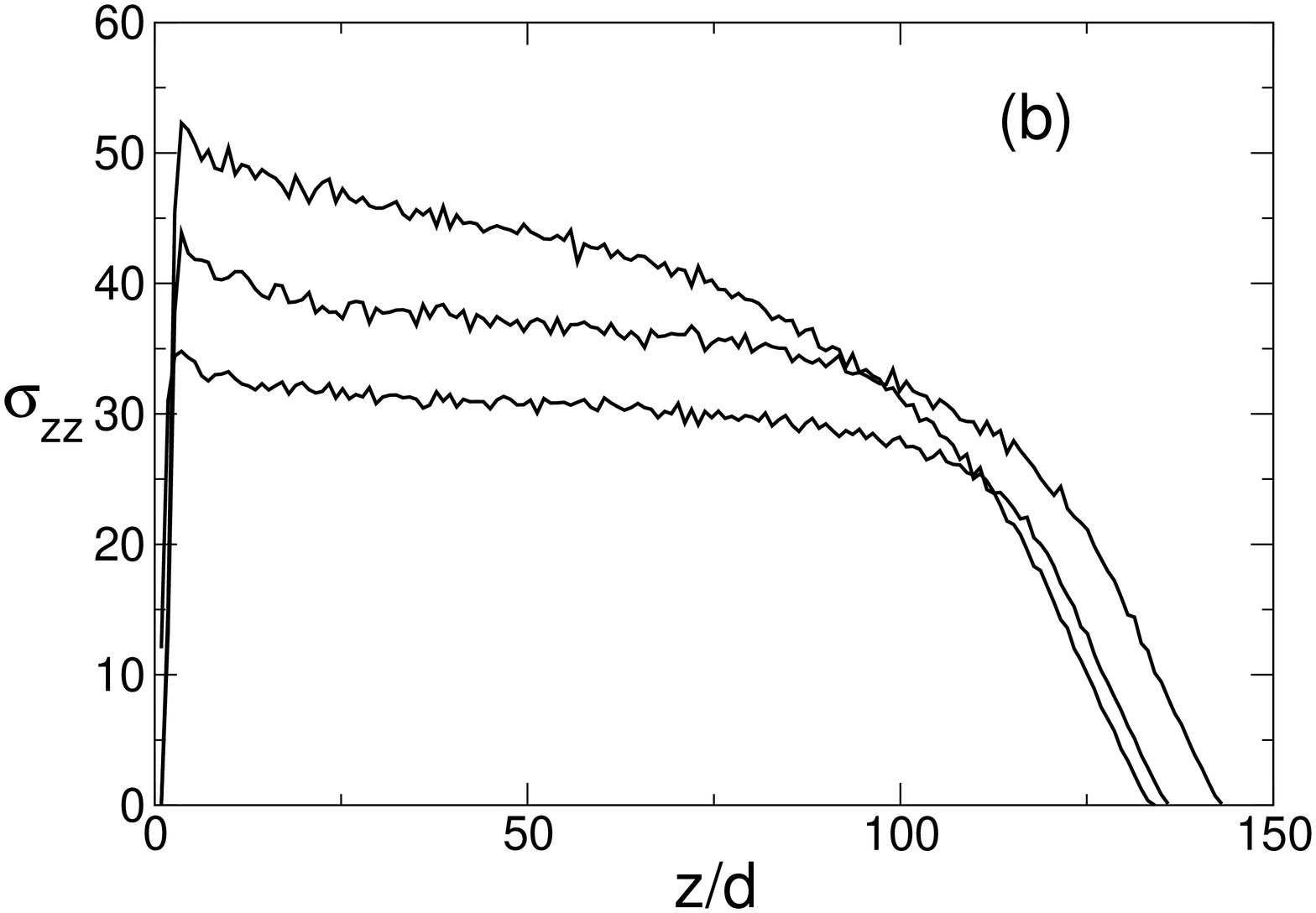}
\caption{\label{fig:stress-diffrad} Comparison of the resultant stress
for packings created with cylinders of different radii $R$.  (a) For
sedimentation method S2, the highest stress is for a $R=20d$ cylindrical
packing with $N=144000$ particles, the second highest is for a $R=15d$
cylindrical packing with 82000 particles, and the lowest is for 50000
particles and $R=10d$.  (b) For pouring method P1, the highest stress
is for a $R=20d$ packing with $N=200000$ particles, next highest is
for a $R=15d$ packing with $120000$ particles, and the lowest is for a
$R=10d$ packing with $50000$ particles.  All the results are a single
run except the two $50,000$ particle systems, which are averaged over 6
runs.}
\end{figure}

We also analyzed stress profiles in larger cylinders of radius $R=15d$
and $R=20d$.  A comparison of different stress profiles is shown in
Figure~\ref{fig:stress-diffrad}a for method S2 and in
Figure~\ref{fig:stress-diffrad}b for method P1.  The wider cylinders
have larger stresses in their asymptotic region because the amount of
material they must support is larger.  These profiles show that the
crossover to height-independent pressure occurs approximately at height
$\approx 6R$, irrespective of pouring method.  In all cases, note the
linear, hydrostatic-like stress region at the top of the pile.

Methods P1 and P2 had similar stress profiles.  Pouring the particles
from different heights had a small effect on the stress profiles.
Increasing the height from which the particles were poured increased the
internal stress.  This arises from their higher potential energy.  The
increase in stress is much greater than the small difference in packing
fraction observed between these packings.  We also varied the pouring
rate for these packings and found this had little or no effect on the
stress profiles.  This leads us to conclude that internal stress in a
packing is primarily affected by the particle-wall friction coefficient
$\mu_w$, the geometry of the cylinder, and the amount of potential
energy that the particles possess, here represented by height of
pouring.  Changes in other parameters that can affect characteristics of
the pack such as packing fraction but do not change the potential energy
have little effect on the stress profiles.


\section{\label{sec:forces}Distribution of Forces}

\begin{figure}
\includegraphics[width=2.25in,angle=270,clip]{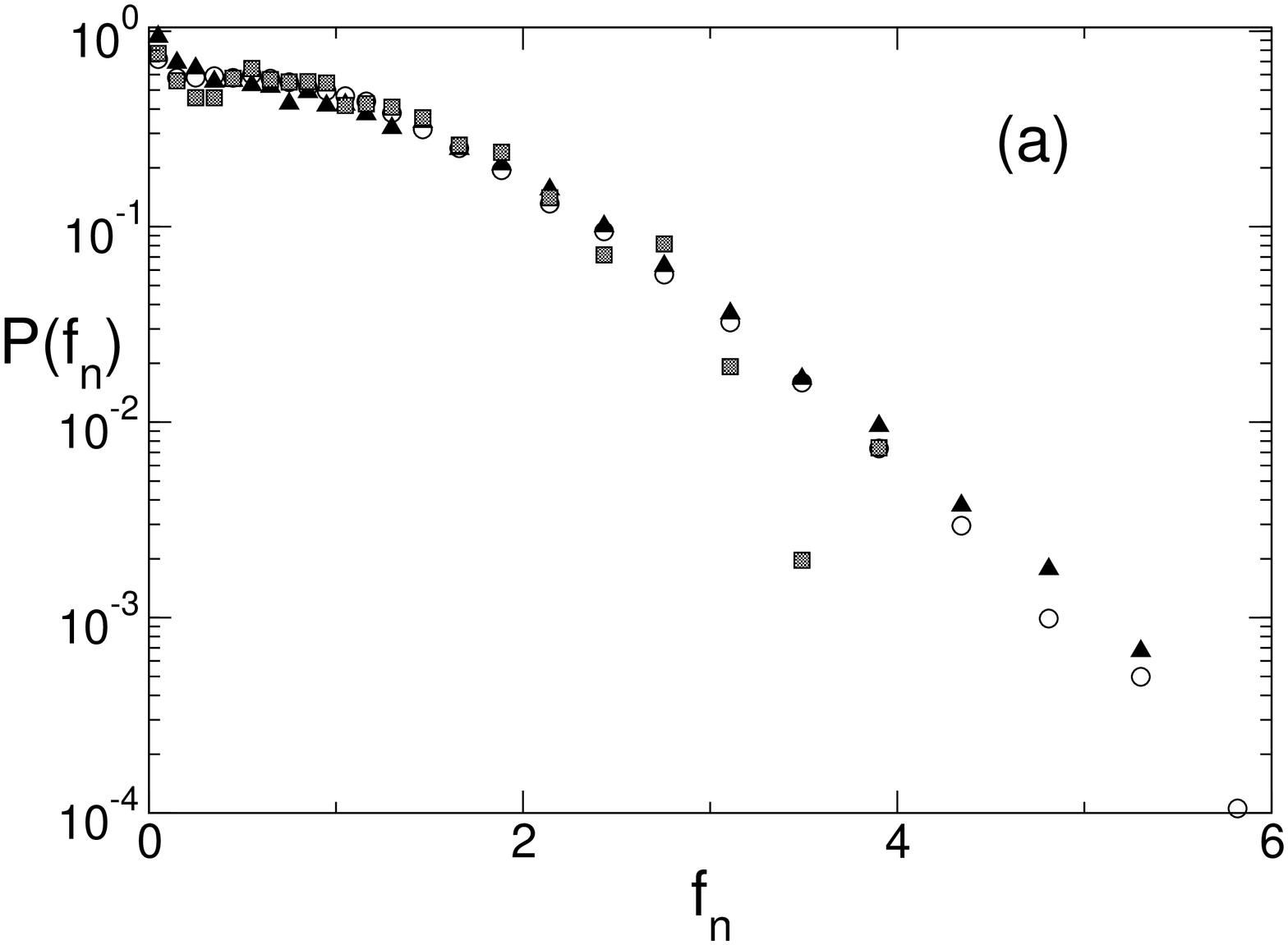}
\includegraphics[width=2.25in,angle=270,clip]{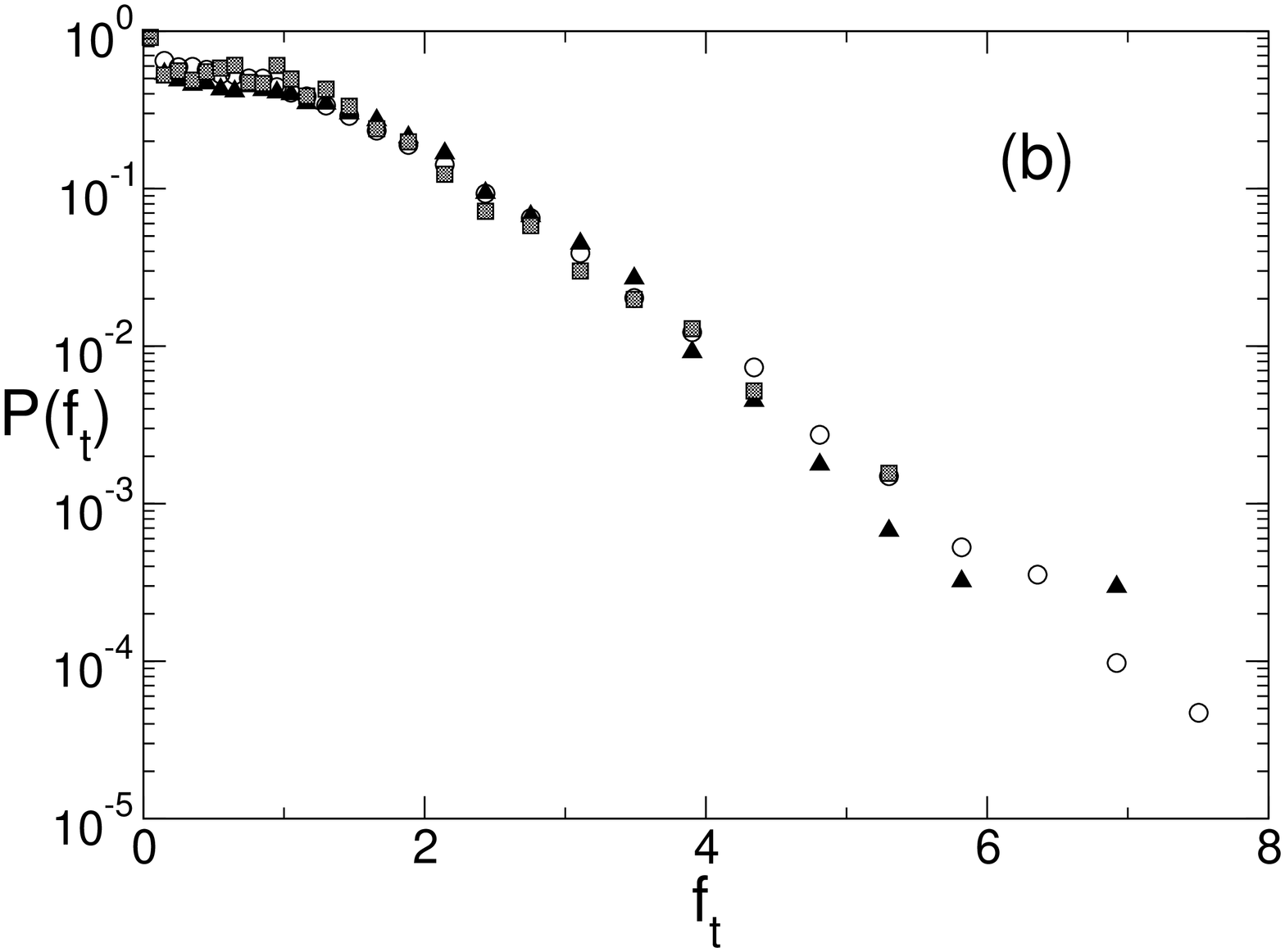}
\caption{\label{fig:pof}Distribution of normal $f_n$ and tangential
$f_t$ contact forces for a packing of $50,000$ particles generated using
method P1.  Bulk forces are represented as open circles, forces between
particles and side wall are represented as filled-in triangles, and
forces between particles and the flat base are represented as filled-in
squares.  All forces exhibit the same quasi-exponential tails.}
\end{figure}

Numerous experiments have been done to measure the distribution of
normal contact forces $P(f_n)$ in granular packings, where $f_n=F_n/\bar
F_n$ and $\bar F_n$ is the average normal force.  These packings all
show approximately exponential tails in $P(f_n)$ for large forces $f_n >
1$~\cite{BlairApr2001,Snoeijer0204277}.  Unfortunately, in experiments
it is difficult to probe the distribution of forces in the interior of
the pack.  We measure $P(f_n)$ in both the bulk of packings and along
the side walls and flat bottoms of the cylinder, shown in
Figure~\ref{fig:pof}.  These packings were created using method P1 with
$\mu_w = 0.5$, though the form of the tail of $P(f)$ is remarkably
robust to changes in method or parameters.  In addition, we see the same
form of the distribution for the tangential $P(f_t)$, as reported in
simulations with periodic packings~\cite{Silbert0208462}.  These
$P(f_n)$ curves are quite consistent with previous measurements of
$P(f_n)$~\cite{BlairApr2001} at the base of a packing.  In addition,
these results indicate the form of $P(f)$ inside a packing is not
qualitatively different from one taken on the edge or bottom of a
cylinder.  Recent experiments on emulsions have found similar
distributions for $P(f)$ in the bulk~\cite{BrujicJun2002,Brujic0210136}.
This implies that measurements of $P(f)$ taken by experiment using
forces at the edge give a good picture of the distribution in the
packing as a whole.

\begin{figure}
\includegraphics[width=2.25in,angle=270,clip]{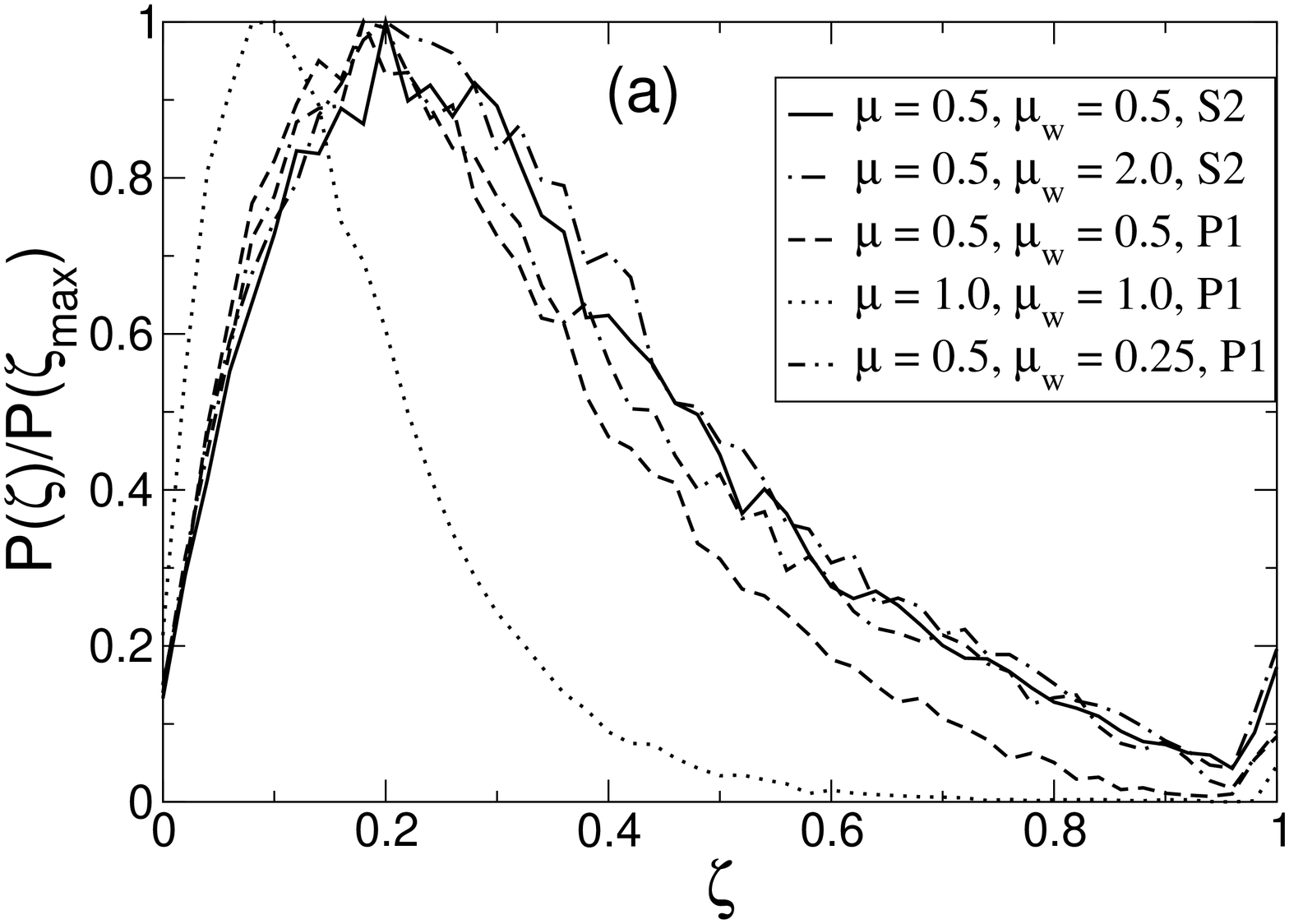}
\includegraphics[width=2.25in,angle=270,clip]{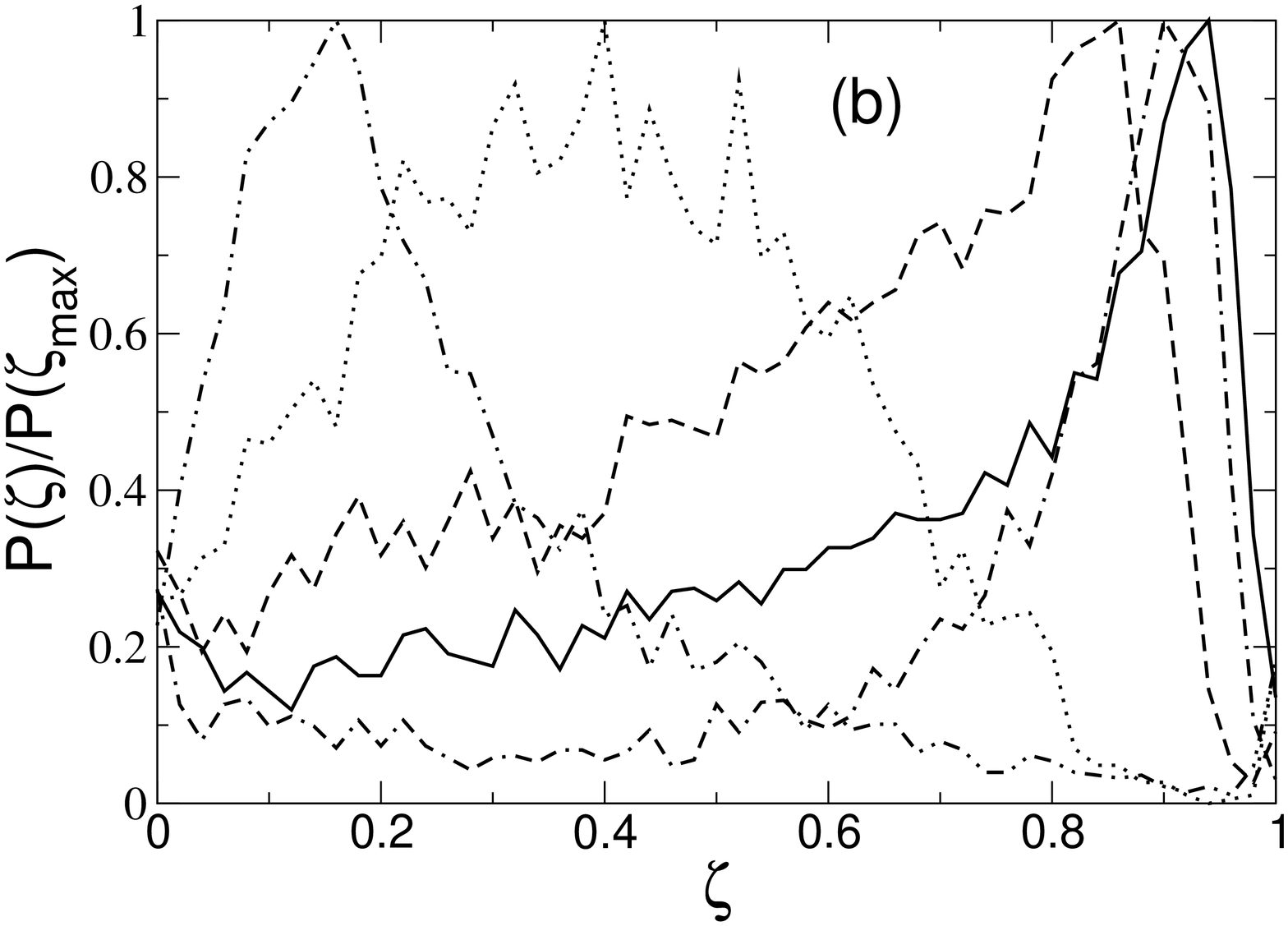}
\caption{\label{fig:coulomb}Probability distributions $P(\zeta)$ in the
height-independent pressure region in the bulk of the packing (a) and at
the side walls (b), each normalized by its maximum value
$P(\zeta_{\mathrm{max}})$.  $\zeta = F_t/\mu F_n$ in (a) and $F_t/\mu_w
F_n$ in (b).  Forces in the bulk are far from the Coulomb failure
criterion, while many of those at the walls are very close to it. The
legends for (a) and (b) are the same.}
\end{figure}

Using our force measurements, we can further test the reliability of the
Janssen assumptions by checking whether the tangential forces at the
wall are actually at the Coulomb yield criterion $F_t = \mu_w F_n$.  We
define $\zeta = F_t/\mu F_n$ in the bulk of the packing and $\zeta =
F_t/\mu_w F_n$ for forces at the wall.  If a specific force is at the
Coulomb failure criterion, $\zeta = 1$.  By examining the distribution
of forces in the interior of our packings, we find that almost no
particle-particle contacts are at the Coulomb criterion irrespective of
method or parameters, as shown in Figure~\ref{fig:coulomb}a.  When we
examine the particle-wall forces in the height-independent stress
region, the forces are much closer to the Coulomb criterion.  For $\mu =
\mu_w = 0.5$, the majority of the tangential forces are close to the
Coulomb criterion for different methods.  When $\mu > \mu_w$, we find
that most of the particle-wall tangential forces are also near the
Coulomb failure criterion.  However, for extremely high-friction walls
($\mu = 0.5$, $\mu_w = 2.0$), most tangential forces are not at the
Coulomb criterion, as shown in Figure~\ref{fig:coulomb}b.  The peak in
the particle-wall distribution occurs near $F_t = \mu F_n$.  This
suggests that there is an effective $\mu_{w,eff}$, which is the lesser
of the original $\mu_w$ and $\mu$.  If we redo the modified Janssen fit
as before for the $\mu_w = 2.0$ case and use an effective $\mu_{w,eff} =
0.5$, as determined from our contact forces, we obtain $\kappa = 0.72$,
a value close to our previous value for $\mu_w = 0.5$, which is what one
would expect.  It appears that the wall does not support in meaningful
numbers larger tangential forces than those between particles, because
particles slip and move against other particles and thus detach from the
wall regardless of the high $\mu_w$.  This suggests that when the
particle-particle friction $\mu$ and particle-wall friction $\mu_w$ are
matched, the majority of the particle-wall forces at the wall are close
to the Coulomb failure criterion.  One exception occurs for large $\mu$,
$\mu = \mu_w = 1.0$.  This allows very large frictional forces, and it
seems likely (as observed in other simulations~\cite{SilbertMar2002})
that even though the wall and particles can support larger tangential
forces in principle, no tangential forces of this magnitude are
generated.  This information about the Coulomb failure criterion in the
depth-independent pressure region gives us no information on the
extended hydrostatic-like region at the top of the pile.

\begin{figure}
\includegraphics[width=2.25in,angle=270,clip]{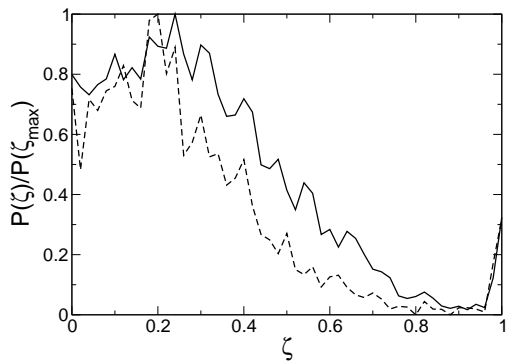}
\caption{\label{fig:coulomb-lin}Probability distributions $P(\zeta)$ at
the side wall in the linear hydrostatic region at the top of the packing
with $\mu = \mu_w = 0.5$, each normalized by its maximum value
$P(\zeta_{\mathrm{max}})$.  $\zeta = F_t/\mu_w F_n$.  The solid line is
the data for method S2 and the dashed line the data for method P1.  In
contrast to the behavior in the height-independent pressure region, the
forces at the walls are far from the Coulomb failure criterion in all
cases.}
\end{figure}

We have also analyzed the linear hydrostatic region specifically and
show our results in Figure~\ref{fig:coulomb-lin}, using $\zeta =
F_t/\mu_w F_n$ as in the earlier figures.  In this region, few of the
forces at the wall are near the Coulomb criteria, regardless of the
value of $\mu$ and $\mu_w$.  This is a partial explanation for why the
Janssen analysis does not apply in this region.  The walls in this
region support very little weight and thus the stress profile in this
region is similar to the linear hydrostatic case.  The nature of the
transition between this hydrostatic-like region and the bulk region
remains to be explored.

\section{\label{sec:conclusion}Conclusions}

We have used large-scale simulations to study granular packings in
cylindrical containers.  We used a variety of methods to generate these
packings and studied the effects of packing preparation on the final
static packing.  We show that the classical Janssen analysis does not
fully describe our packings, but that slight modifications to the theory
of Janssen enable us to describe our packings well.  In addition, we explore
some of the assumptions of Janssen and show that when the
particle-particle and particle-wall friction interactions are balanced,
the particle-wall interaction close to the wall is at the Coulomb
failure criterion.  We show that the anomalous hydrostatic region at the
top of our packings arises because the forces at the wall are far from
the Coulomb failure criterion and thus support very little weight, in
contrast to results deeper in the packing.  We also demonstrate that the
distribution of forces in our packings is consistent with previous
results in both experiment and simulation not only in the bulk, but also
at the walls and base.

Much of the literature on vertical stress profiles in silos focuses on
two dimensional systems.  The stress profiles of packings are strongly
influenced by the dimensionality of the system and we explore the
crossover between $2D$ packings, quasi-$2D$ packings of particles in
flat cells, and fully $3D$ packings in another
work~\cite{LandryNov2002}.

While this work was being prepared, we became aware of two new granular
experiments that find a Janssen form for the vertical
stress~\cite{Ovarlez0212228, Bertho0211510}.  These experiments use
either a movable base or movable cylindrical walls to mobilize the
grains more fully, producing a more Janssen-like vertical stress.  The
present simulations are much closer to the Vanel-Cl\'{e}ment
experiments~\cite{VanelMay1999,VanelFeb2000}, which were taken after the
packing had settled into its final state.

This work was supported by the Division of Materials Science and
Engineering, Basic Energy Sciences, Office of Science, U.S. Department
of Energy.  This collaboration was performed under the auspices of the
DOE Center of Excellence for the Synthesis and Processing of Advanced
Materials.  Sandia is a multiprogram laboratory operated by Sandia
Corporation, a Lockheed Martin Company, for the United States Department
of Energy under Contract DE-AC04-94AL85000.

\bibliography{grain}

\begin{thebibliography}{37}
\bibitem[{Pac()}]{PackingNote1}
\bibinfo{note}{In the Hookean case, energy drains out of the pile efficiently
  and additional damping is not required. For Hertzian contacts, tangential
  damping is required to allow the energy in the pile to dissipate in a
  reasonable computational time.}
\bibitem[{Ras()}]{Raster}
\bibinfo{note}{These figures were created using the raster3d program version
  2.6e.}
\end{thebibliography}

\end{document}